\documentclass{kapproc} 

\RequirePackage{graphicx}%
\RequirePackage{epsf}%
\input{psfig.sty}

\upperandlowercase
\let\footnote\savefootnote
\let\footnotetext\savefootnotetext 
 
\kluwerbib 

\def\msun{$M_\odot$\ }
\def\msunp{$M_\odot$}

\begin{document}


\articletitle{Embedded Clusters and the IMF}


\chaptitlerunninghead{Embedded Cluster IMFs}



 \author{Charles J. Lada}
 \affil{Smithsonian Astrophysical Observatory, 60 Garden St., Cambridge, MA USA}
 \email{clada@cfa.harvard.edu}





 \begin{abstract}
Despite valiant efforts over nearly five decades, attempts to determine the IMF
over a complete mass range for galactic field stars and in open clusters have
proved difficult.  Infrared imaging observations of extremely young embedded
clusters coupled with Monte Carlo modeling of their luminosity functions are
improving this situation and providing important new contributions to our
fundamental knowledge of the IMF and its universality in both space and time.

 \end{abstract}

\section{Introduction}

A fundamental consequence of the theory of stellar structure and evolution is that,
once formed, the subsequent life history of a star is essentially predetermined by
one parameter, its birth mass.  Consequently, detailed knowledge of the initial
distribution of stellar masses at birth (i.e., the IMF) and how this quantity
varies through time and space is necessary to predict and understand the evolution
of stellar systems, such as galaxies and clusters.  Unfortunately, stellar
evolution theory is unable to predict the form of the IMF.  This quantity must be
derived from observations.
Stellar clusters have played an important role in IMF studies because they present
equidistant and coeval populations of stars of similar chemical composition.
Compared to the disk population, clusters provide an instantaneous sampling of the
IMF at different epochs in galactic history (corresponding to the different cluster
ages) and in different, relatively small volumes of space.  This enables
investigation of possible spatial and temporal variations in the IMF.  Extremely
young embedded clusters are particularly useful laboratories for IMF measurements
because these clusters are too young to have lost significant numbers of stars due
to stellar evolution or dynamical evaporation, thus their present day mass
functions are, to a very good approximation, their initial mass functions.
Embedded clusters are also particularly well suited for determining the nature of
the IMF for low mass stars and substellar objects.  This is because low mass
stars in embedded clusters are primarily pre-main sequence stars, and thus are
brighter than at any other time in their lives. At these early ages brown dwarfs
are similarly bright as low mass stars. Indeed, infrared observations of
modest depth are capable of detecting objects spanning the entire range of stellar
mass from 0.01 to 100 \msun in clusters within 0.5 -- 1.0 Kpc of the sun.

\section{From Luminosity to Mass Functions}

The monochromatic brightness of a star is its most basic observable property and
infrared cameras enable the simultaneous measurement of the infrared monochromatic
brightnesses of hundreds of stars.  Thus, complete luminosity functions, which span
the entire range of stellar mass, can be readily constructed for embedded stellar
clusters with small investments of telescope time.  The monochromatic (e.g., K
band) luminosity function of a cluster, $dN \over dm_K$, is defined as the number
of cluster stars per unit magnitude interval and is the product of the underlying
mass function and the derivative of the appropriate mass-luminosity relation (MLR):

\begin{equation} 
{dN\over dm_K} = {dN\over dlogM_*} \times {dlogM_*\over dm_K} 
\label{eq1} 
\end{equation}


\noindent
where $m_k$ is the apparent stellar (K) magnitude, and $M_*$ is the stellar mass.  The
first term on the right hand side of the equation is the underlying stellar mass
function and the second term the derivative of the MLR.  With knowledge of the MLR (and
bolometric corrections) this equation can be inverted to derive the underlying mass
function from the observed luminosity function of a cluster whose distance is known.

This method is essentially that originally employed by Salpeter (1955) to derive
the field star IMF.  However, unlike main sequence field stars, PMS stars, which
account for most of the stars in the an embedded cluster, cannot be characterized
by a unique MLR.  Indeed, the MLR for PMS stars is a function of time.  Moreover,
for embedded clusters the duration of star formation can be a significant fraction
of the cluster's age.  Consequently, to invert the equation and derive the mass
function one must model the luminosity function of the cluster and this requires
knowledge of both the star formation history (i.e., age and age spread) of the
cluster as well as the time-varying PMS mass-luminosity relation.  The
age or star formation history of the cluster typically can be
derived by placing cluster stars on an HR diagram.  This, in turn, requires
additional observations such as multi-wavelength photometry or spectroscopy of a
representative sample of the cluster members. PMS models must be employed
to determine the time varying mass-luminosity relation.  The accuracy of the
derived IMF therefore directly depends on the accuracy of the adopted PMS models
which may be inherently uncertain, particularly for the youngest clusters ($\tau <
10^6$ yrs) and lowest mass objects ($m < 0.08$ \msun).  

Despite these complexities, Monte Carlo modeling of the infrared luminosity
functions of young clusters (Muench, Lada \& Lada 2000) has demonstrated that {\it
the functional form of an embedded cluster's luminosity function is considerably
more sensitive to the form of the underlying cluster mass function than to any
other significant parameter} (i.e., stellar age distribution, PMS models, etc.).
In particular, despite the significant differences between the parameters that
characterize the various PMS calculations (e.g., adopted convection model,
opacities, etc.), model luminosity functions were found to be essentially
insensitive to the choice of existing PMS mass-to-luminosity relations.  This
indicated that, given smoothly varying mass-luminosity relations and knowledge of
their ages, the monochromatic luminosities of PMS stars can provide very good
proxys for their masses.  {\it This is a direct result of the steepness of the
mass-luminosity relation for PMS stars.}

The top panel of Figure 1 shows the K luminosities (magnitudes) for
million-year-old PMS stars predicted by a suite of the best known PMS models in the
literature.  The excellent agreement between the various models reflects the steep
dependence of luminosity on stellar mass, a consequence of the basic stellar
physics of Kelvin-Helmholtz contraction.  Any intrinsic variations or uncertainties
in the models are dwarfed by the sensitivity of luminosity to stellar mass.  This
is in contrast to the situation for the predicted stellar effective temperatures as a
function of mass.  The bottom panel of Figure 1 shows that the predicted effective
temperatures are much less sensitive to stellar mass.  The intrinsic variations in
the models are roughly similar in magnitude to the overall variation in effective
temperature with mass.

\begin{figure}[ht]
\vskip -0.8in
\centerline{\psfig{file=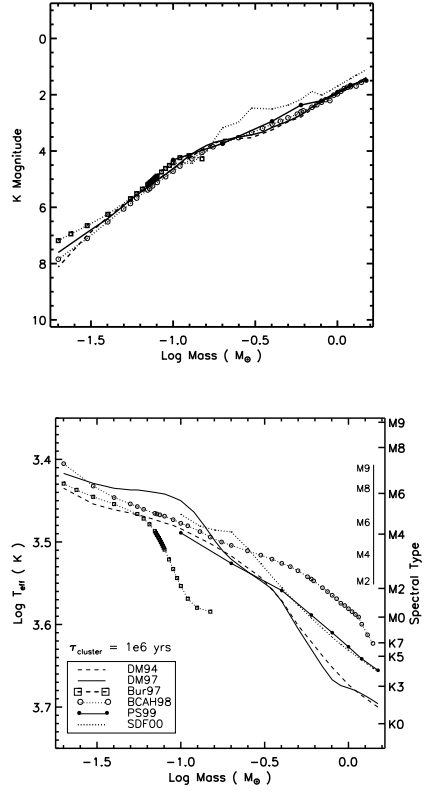,height=7in}}
\vskip -2in
\caption{Comparisons of theoretical predictions for the infrared luminosities and
effective temperatures of million-year-old PMS stars as a function of
mass from a suite of standard PMS models (Burrows et al. 1997; Baraffe
et al. 1998; D'Antona \& Mazzitelli 1994; 1997; Palla \& Sthaler 1999;
Seiss et al. 2000). Note that the predicted PMS K magnitudes (top)
appear to be in excellent agreement across the entire mass range whereas 
the predicted PMS effective temperatures (bottom) are not. This is a result 
of the steepness of the infrared mass-luminosity relation and clearly demonstrates
how sensitive PMS luminosity is to variations in stellar mass. }
\end{figure}

\begin{figure}[ht]
\centerline{\psfig{file=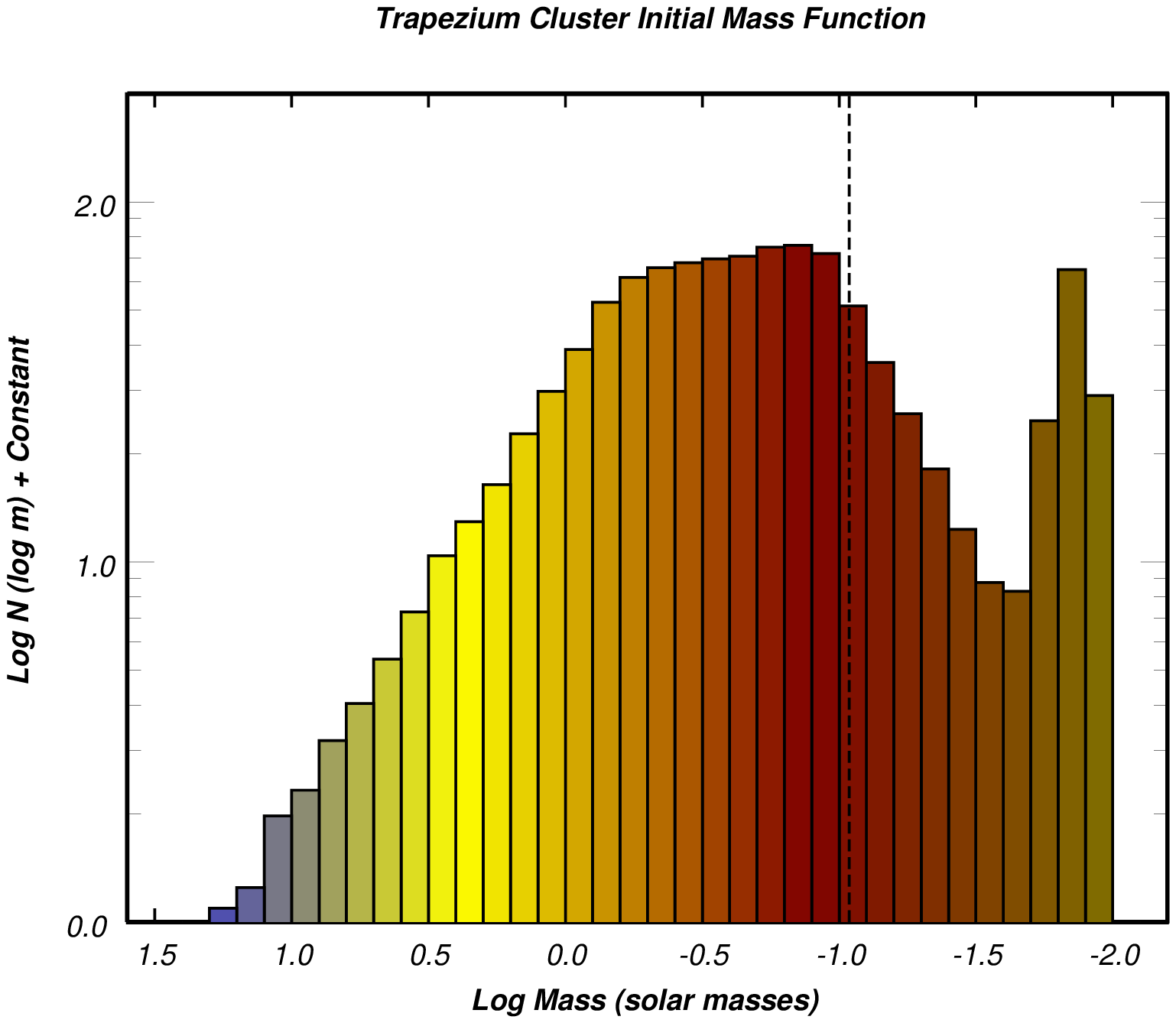,height=3in}}
\caption{The IMF derived for the Trapezium cluster from Monte Carlo modeling
of its luminosity function (Muench et al. 2002). This plot displays the binned
mass function of the synthetic cluster whose luminosity function was found to
best fit the observed KLF of the Trapezium cluster. A vertical dashed line
marks the approximate location of the hydrogen burning limit. The derived IMF
extends deep into the substellar regime.}
\end{figure}

\section{The IMF from OB stars to Brown Dwarfs}

The rich Trapezium cluster in Orion represents the best nearby target for
determining the IMF of a young stellar population.  Muench et al.  (2002) obtained
deep infrared images of the Trapezium cluster and derived its IMF by using a suite
of Monte Carlo calculations to model the cluster's K-band luminosity function
(KLF).  The observed shape of a cluster luminosity function depends on three
parameters:  the ages of the cluster stars, the cluster mass-luminosity relation,
and the underlying IMF (i.e., Equation~\ref{eq1}).  With the assumptions of a fixed
age distribution, derived from an existing spectroscopic study of the cluster by
Hillenbrand (1997), a composite theoretical mass-luminosity relation adopted from
published PMS calculations and an empirical set of bolometric corrections, Muench
et al.  varied the functional form of the underlying IMF to construct a series of
synthetic KLFs.  These synthetic KLFs were then compared to the observed,
background corrected, Trapezium KLF in a Chi-Squared minimization procedure to
produce a best-fit IMF.  As part of the modeling procedure, the synthetic KLFs were
statistically corrected for both variable extinction and infrared excess using
Monte Carlo probability functions for these quantities that were derived directly
from multi-color observations of the cluster.

The derived mass function is displayed in Figure~2 in the form of a histogram of
binned masses of the stars in the best-fit synthetic cluster.  This model mass
function represents the IMF of the young Trapezium cluster.  This mass function
agrees very well with Trapezium IMFs derived from a number of other different deep
infrared imaging surveys using a variety of methods (Lucas \& Roche 2000;
Hillenbrand and Carpenter 2000; Muench et al.  2000; Luhman et al.  2000).  The
main characteristics of this IMF are:  1) the sharp power-law rise of the IMF from
about 10 \msun (OB stars) to 0.6 \msun (dwarf stars) with a slope (i.e., $\beta$ =
--1.2) similar to that of Salpeter (1955), 2) the break from the single power-law
rise at 0.6 \msun followed by a flattening and slow rise reaching a peak at about
0.1 \msunp, near the hydrogen burning limit, and 3) the immediate steep decline
into the substellar or brown dwarf regime.

The most significant characteristic of this IMF is the broad peak, extending
roughly from 0.6 to 0.1 \msunp.   This structure clearly demonstrates that
{\bf there is a characteristic mass produced by the star formation process in Orion.}
That is, the typical outcome of the star formation process in this cluster is a
star with a mass between 0.1 and 0.6 \msunp.  The process produces relatively few
high mass stars and relatively few substellar objects.  {\it Indeed, no more than
$\sim$ 22\% of all the objects formed in the cluster are freely floating brown
dwarfs}.  The overall continuity of the IMF from OB stars to low mass stars and
across the hydrogen burning limit strongly suggests that the star formation process
has no knowledge of the physics of hydrogen burning.  Substellar objects are
produced naturally as part of the same physical process that produces OB stars.

The derived IMF of the Trapezuim cluster spans a significantly greater range of
mass than any previous IMF determination whether for field stars or other clusters
(e.g., Kroupa 2002).  Its statistically meaningful extension to substellar masses
and the clear demonstration of a turnover near the HBL represents an important
advance in IMF studies.  For masses in excess of the HBL the IMF for the Trapezium
is in good agreement with the most recent determinations for field stars (Kroupa
2002).  This is to some extent both remarkable and surprising since the field star
IMF is averaged over billions of years of galactic history, assuming a constant
star formation rate, and over stars originating from very different locations of
galactic space.  The Trapezium cluster, on the other hand, was formed within the
last million years in a region considerably less than a parsec in extent.  Taken at
face value this agreement suggests that the IMF and the star formation process that
produces it are very robust in the disk of the Galaxy.

%

\end{document}